\pdfoutput=1
\documentclass[final]{article}

\usepackage{times}
\usepackage{ifpdf}
\makeatletter
\def\@normalsize{\@setsize\normalsize{12pt}\xpt\@xpt
\abovedisplayskip 10pt plus2pt minus5pt\belowdisplayskip \abovedisplayskip
\abovedisplayshortskip \z@ plus3pt\belowdisplayshortskip 6pt plus3pt
minus3pt\let\@listi\@listI}

\def\section{\@startsection {section}{1}{\z@}{10pt plus 2pt minus 2pt}
{8pt plus 2pt minus 2pt}{\centering\normalsize\sc
\edef\@svsec{\thesection.\ }}}
\def\thesection{\Roman{section}}

\def\subsection{\@startsection {subsection}{2}{\z@}{10pt plus 2pt minus 2pt}
{6pt plus 2pt minus 2pt}{\normalsize\sl
\edef\@svsec{\thesubsection.\ }}}
\def\thesubsection{\Alph{subsection}}

\long\def\@makecaption#1#2{
\vskip6pt\begin{center} #1 #2 \end{center}\par\vskip 1pt}
\def\fnum@figure{\raggedright{\footnotesize Fig. \thefigure }.%
\footnotesize}
\def\fnum@table{\footnotesize TABLE \thetable\\\footnotesize\sc}
\def\thetable{\Roman{table}}

\partopsep          \z@

\parsep             \z@

\itemsep            \z@

\labelwidth         \z@

\def\@listi{\leftmargin\leftmargini \topsep 2pt plus 1pt minus 1pt}
\let\@listI\@listi
\def\@listii{\leftmargin\leftmarginii\labelwidth\leftmarginii%
    \advance\labelwidth-\labelsep \topsep 2pt}
\def\@listiii{\leftmargin\leftmarginiii\labelwidth\leftmarginiii%
    \advance\labelwidth-\labelsep \topsep 2pt}
\def\@listiv{\leftmargin\leftmarginiv\labelwidth\leftmarginiv%
    \advance\labelwidth-\labelsep \topsep 2pt}
\def\@listv{\leftmargin\leftmarginv\labelwidth\leftmarginv%
    \advance\labelwidth-\labelsep \topsep 2pt}
\def\@listvi{\leftmargin\leftmarginvi\labelwidth\leftmarginvi%
    \advance\labelwidth-\labelsep \topsep 2pt}

\makeatother

\usepackage{graphicx,cite}
\usepackage{url}
\usepackage{booktabs}
\usepackage{graphicx}
\usepackage{amssymb}
\usepackage{amsmath}
\usepackage{wasysym}

\begin{document}
%
\title{\Large\bf Rule-Based Optimization of Reversible Circuits}
\date{}

%
\author{\normalsize
 \begin{tabular}[]{l c r}
   \hspace{27 pt}\large Mona Arabzadeh & \hspace{8 pt} \large Mehdi Saeedi &  \hspace{8 pt} \large Morteza Saheb Zamani \\
  \\
\end{tabular}
\\
   \normalsize Computer Engineering and IT Department \\
  \normalsize Amirkabir University of Technology\\
  \normalsize Tehran, Iran, PO BOX: 15875-4413\\
  \normalsize E-mail: \{mona\_arabzadeh, msaeedi, szamani\}@aut.ac.ir\\
}
\maketitle
\thispagestyle{empty}

\newcounter {TCounter}
\newcounter {LCounter}
\newcounter {PCounter}
\newcounter {ACounter}
\newcounter {CCounter}
\newcounter {DCounter}
\newcounter {ECounter}

\newtheorem{theorem}[TCounter]{\textbf{Theorem}}
\newtheorem{lemma}[LCounter]{\textbf{Lemma}}
\newtheorem{proposition}[PCounter]{\textbf{Proposition}}
\newtheorem{algorithm}[ACounter]{\textbf{Method}}
\newtheorem{corollary}[CCounter]{\textbf{Corollary}}
\newtheorem{definition}[DCounter]{\textbf{Definition}}
\newtheorem{example}[ECounter]{\textbf{Example}}

\newenvironment{proofs}[1][\textbf{Proof}]{\begin{trivlist}
\item[\hskip \labelsep {\bfseries #1}]}{\end{trivlist}}
\newenvironment{remark}[1][\textbf{Remark}]{\begin{trivlist}Alg:1
\item[\hskip \labelsep {\bfseries #1}]}{\end{trivlist}}

\newcommand{\qed}{\nobreak \ifvmode \relax \else
      \ifdim\lastskip<1.5em \hskip-\lastskip
      \hskip1.5em plus0em minus0.5em \fi \nobreak
      \vrule height0.75em width0.5em depth0.25em\fi}

{\small\bf Abstract---
Reversible logic has applications in various research areas including low-power design and quantum computation. In this paper, a rule-based optimization approach for reversible circuits is proposed which uses both negative and positive control Toffoli gates during the optimization. To this end, a set of rules for removing NOT gates and optimizing sub-circuits with common-target gates are proposed. To evaluate the proposed approach, the best-reported synthesized circuits and the results of a recent synthesis algorithm which uses both negative and positive controls are used. Our experiments reveal the potential of the proposed approach in optimizing synthesized circuits.}

\section {Introduction}
A Boolean function is called reversible if it maps each input assignment to a unique output assignment. Landauer \cite{Landauer61} proved that using conventional irreversible logic gates leads to energy dissipation, regardless of the underlying circuit. Today, reversible logic has received considerable attention in various research areas including low-power CMOS design \cite{Schrom98} and quantum computing \cite{Nielsen00}.

Reversible logic synthesis deals with generating an efficient reversible circuit from a given reversible specification. The synthesis of reversible circuits differs from that of traditional irreversible ones with respect to the characterizations of reversible logic. For examples, loop and fanout are not allowed in reversible logic. Therefore, the available synthesis algorithms for irreversible circuits cannot be applied to reversible specifications directly.

In order to generate an efficient reversible circuit from a given specification, different scenarios have been applied in the recent years. Among them, a two-step synthesis and optimization approach has been widely used recently. In this scenario, a realization is found from a given specification first and then, further optimizations are applied in a post-processing step to improve various cost metrics (e.g., quantum cost). Local transformation \cite{IwamaDAC02}, templates matching \cite{MillerDAC03,MaslovTODAES07, MaslovDATE05, MDM:2005}, and data-structure-based optimization \cite{Prasad06} were used to simplify synthesized circuits in the past.

In this paper, we propose a rule-based optimization approach for reversible circuits to improve the quantum cost. To this end, multiple control Toffoli gates with both positive and negative controls are used. While the potential advantage of using negative control Toffoli gates for the simplification of reversible circuits has been announced before (e.g., see \cite{MaslovTCAD08}), this is the first attempt to use negative control Toffoli gates for improving the quantum cost of the synthesized circuits.

The remainder of the paper is organized as follows. In Section \ref{sec:basic_concepts} basic concepts are explained. Previous work is discussed in Section \ref{sec:previous_work}. Our rule-based optimization approach is proposed in Section \ref{sec:prop_method}. Experimental results are reported in Section \ref{sec:exper_res} and finally, Section \ref{sec:conc} concludes the paper.

\section{Preliminaries} \label{sec:basic_concepts}
An $n$-input, $n$-output, fully specified Boolean function $f:\mathbb{B}^n\rightarrow \mathbb{B}^n$ over variables $X=\{x_1,\dots , x_n\}$ is called \emph{reversible} if it maps each input pattern to a unique output pattern. In this paper, $n$ is particularly used to refer to the number of inputs/outputs of a circuit (circuit size). Outputs that are not required in the function specification are considered as \emph{garbage} or \emph{auxiliary} bits. An $n$-input, $n$-output \emph{gate} is reversible if it realizes a reversible function. Previously, various reversible gates with different functionalities have been proposed. Among them, multiple control Toffoli gate has been used by different synthesis methods (e.g., see \cite{ MaslovTODAES07, SaeediICCAD07, WilleDAC09, Gupta06, MDM:2005, SaeediISVLSI07, WGDD:2009, MaslovTCAD08}) and is defined as:

A \emph{multiple control Toffoli gate} C$^m$NOT can be written as C$^m$NOT(C; t), where $C = \{x_{i_1} , \dots , x_{i_m}\} \subset X$ is the set of \emph{control} lines and $t = \{x_j\}$ with $C \cap t = \emptyset$ is the \emph{target} line. The value of the target line is inverted iff all control lines have the required zero or one values. A control line may be \emph{positive} (\emph{negative}) which means that if its value is one (zero), the target is inverted. For $m$=0 and $m$=1, the gates are called NOT and CNOT, respectively. For $m$=2, the gate is called C$^2$NOT or Toffoli.

In addition to the C$^m$NOT gate, several other gate types have been proposed in the literature (see \cite{Nielsen00} for some examples). Controlled-V (controlled-V$^+$) changes the value on its target line using the transformation given by the matrix V (V$^+$) if the control line has the value of 1.
\[
V = \frac{{1 + i}}{2}\left[ {\begin{array}{*{20}c}
   1 & { - i}  \\
   { - i} & 1  \\
\end{array}} \right],V^ +   = \frac{{1 - i}}{2}\left[ {\begin{array}{*{20}c}
   1 & i  \\
   i & 1  \\
\end{array}} \right]
\]

\begin{figure}[t]
	\centering
		\includegraphics[scale=0.55]{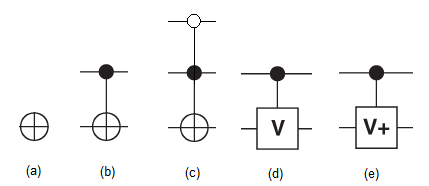}
		\caption{Different gate types, (a) NOT, (b) CNOT, (C) C$^2$NOT, (d) controlled-V, and (e) controlled-V$^+$.}
	\label{Fig:0}
\end{figure}

Fig. \ref{Fig:0} shows different gate types where positive and negative controls are denoted by \CIRCLE \, and \Circle \, symbols, respectively.

The gates NOT, CNOT, controlled-V, and controlled-V$^+$ (with positive controls) have been efficiently simulated in some quantum computer technologies \cite{Lee06}. These gates are considered as \emph{elementary gates} for reversible Boolean functions \cite{Barenco95,MaslovDATE05}. The number of elementary gates required for simulating a given gate is called \emph{quantum cost}. A \emph{reversible circuit} includes a set of reversible gates.

Consider a circuit of size $n$. For ($n \geq 5$), a C$^m$NOT gate ($m \in \left\{3,4, \cdots, \left\lceil n/2 \right\rceil\right\}$) can be simulated by $12m-22$ elementary gates if at least one positive control is available; otherwise, two extra elementary gates are required  \cite{MaslovDATE05, MaslovTCAD08}. In addition, for $n \geq 7$, a C$^{n-2}$NOT gate can be simulated by $24n-88$ elementary gates with no auxiliary bits if there is at least one positive control \cite{MaslovDATE05,MaslovTCAD08}. On the other hand, for a C$^{n-2}$NOT with only negative controls, four additional elementary gates are required \cite{MaslovTCAD08}. Moreover, a C$^{n-1}$NOT gate can be simulated with an exponential cost $2^n-3$ if no garbage line is available and all controls are positive \cite{Barenco95}. A C$^{n-1}$NOT gate with at least one positive control has the same $2^n-3$ cost. For the case of all negative controls, two additional elementary gates should be applied.
%

To avoid the exponential size and the need for a large number of elementary gates, several researchers used an extra garbage line for an efficient simulation of C$^{n-1}$NOT gate (e.g., \cite{MaslovTODAES07}). Generally, the number of available bits is very restricted in today's reversible and quantum implementations \cite{Negrevergne}. Therefore, for two circuits with equal linear costs, the one without garbage line is preferred. Note that a C$^2$NOT gate has the cost of 5 if at least one positive control exists. Otherwise, six elementary gates are required for the optimal implementation \cite{MaslovIET07}. In addition, the quantum cost for a CNOT gate with negative control is 3.

For the purpose of optimization, two adjacent gates can be interchanged if the target of the first gate is not one of the controls of the second gate and vice versa (\emph{moving rule}). In addition, two adjacent gates with the same functionalities can be canceled (\emph{deletion rule}) \cite{MillerDAC03}.

\section{Previous Work} \label{sec:previous_work}

During the recent years, several algorithms have been proposed to address the requirements of the synthesis and optimization steps. In this section, we review optimization-related algorithms. However, as the algorithm of \cite{SaeediICCAD07} is used in our experiments, we briefly explain it first.

A non-search based synthesis algorithm was proposed in \cite{SaeediICCAD07} which works on the truth table columns of a given specification to gradually transform the truth table into the identity function. The algorithm always converges and it leads to a valid result very fast compared with those methods that explore the search space. Multiple control Toffoli gates with both positive and negative lines were used in this method.

In \cite{IwamaDAC02} a set of local transformation rules for reversible circuits was proposed. It was shown that the set was complete which means that for any two equivalent circuits, there is a sequence of transformations which change one of the circuits to the other. This rule set was used in \cite{IwamaDAC02} for developing a design theory for Boolean reversible circuits and improving their cost.

The application of rule set was extended in \cite{MillerDAC03} where the authors introduced several transformation rules based on a set of predefined patterns called \emph{templates}. According to \cite{MillerDAC03}, a template $T$ is a circuit with identity function which contains $m$ gates $g_1$, $g_2$, $\cdots$, $g_m$. Consider the first $k$ ($k > m/2$) gates of $T$ (i.e., $g_1$, $g_2$, $\cdots$ $g_k$) and suppose that these gates are applied in a reversible circuit in sequence. It can be verified that the set of $m-k$ gates $g_m$, $\cdots$, $g_{k+2}$, $g_{k+1}$ can be applied instead of the initial $g_1$, $g_2$, $\cdots$ $g_k$ gates to reduce the gate count from $k$ to $m-k$. A similar method can be applied to reduce the quantum cost. The template-based optimization was used in several papers recently (see \cite{MaslovTODAES07, MDM:2005, MaslovTCAD08}).

The authors of \cite{Prasad06} developed a data structure to generate and store optimal circuits for all reversible functions of size 3 and many of four inputs circuits. Using the proposed representation and algorithm, the authors guaranteed to have optimal implementations for all sub-circuits of size 3 and many of size 4 functions. In a given reversible circuit, a sub-circuit is examined if it contains less than 5 variables. Then, its optimal implementation is explored in a pre-constructed library and is replaced with the initial implementation if the optimal implementation is found.

\section{Proposed Method} \label{sec:prop_method}

In this section, after proposing a transformation rule for moving a NOT gate across a given reversible circuit, an optimization method is proposed to find the optimal realization of a sequence of gates with the same targets. Next, the proposed methods will be used to optimize the results of the available synthesis algorithms in a post-processing step.

\subsection{NOT Reduction}
\begin{figure}[t]
	\centering
		\includegraphics[scale=0.5]{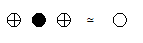}
		\caption{Negative control definition}
	\label{Fig:2}
\end{figure}

\begin{figure}[t]
	\centering
		\includegraphics[scale=0.5]{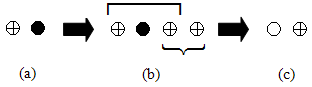}
		\caption{PR, (a) before transformation, (b) negative control definition, (c) after transformation  }
	\label{Fig:3}
\end{figure}

\begin{figure}[t]
	\centering
		\includegraphics[scale=0.5]{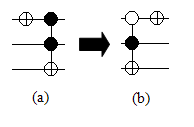}
		\caption{PR transformation, (a) before transformation, (b) after transformation}
	\label{Fig:1}
\end{figure}

By moving NOT gates through a given reversible circuit, one can delete redundant NOTs to improve the total cost according to the following rule:

\begin{definition}
\emph{Pass Rule} (PR): A NOT($x$) gate can be interchanged with its adjacent C$^m$NOT(C;t) gate without changing the functionality of the circuit if at least one of the following conditions are met:
\begin{itemize}
	\item $x \notin C, x \neq t$.
	\item $x \notin C, x=t$.
	\item $x \in C, x \neq t$. For this case, the polarity of the control $x$ of the C$^m$NOT(C;t) gate is toggled. In other words,	a negative control line can be considered as a positive control with two NOTs as shown in Fig. \ref{Fig:2}. Now, consider the controls in Fig. \ref{Fig:3}. Two NOT gates can be inserted after the positive control as shown in Fig. \ref{Fig:3}-b which leads to Fig. \ref{Fig:3}-c. See Fig. \ref{Fig:1} for one example.
\end{itemize}
\end{definition}

%

\begin{figure}[t]
	\centering
		\includegraphics[scale=0.5]{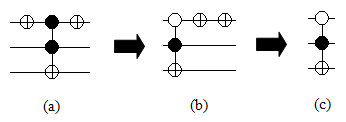}
		\caption{A reversible circuit (a) before PR, (b) after PR, (c) after canceling redundant NOTs}
	\label{Fig:4}
\end{figure}

It is worth noting that by applying PR, an even number of adjacent NOTs can be canceled to reduce the quantum cost. See Fig. \ref{Fig:4} for an example where the cost was improved from 7 to 5.

\begin{figure}[t]
	\centering
		\includegraphics[scale=0.5]{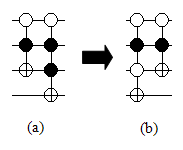}
		\caption{GPR transformation, (a) before transformation, (b) after transformation }
	\label{Fig:5}
\end{figure}

\begin{definition}
\emph{Generalized Pass Rule} (GPR): A C$^{n-1}$NOT($C_1$;$t_1$) gate can be interchanged with its adjacent $C^{n-2}$NOT($C_2$;$t_2$) gate without changing the functionality of the circuit if $C_1={C_2 \cup t_2}$. In this case, the polarity of the control line $t_2$ of C$^{n-1}$NOT($C_1$;$t_1$) is toggled (see Fig. \ref{Fig:5} for an example).
\end{definition}

%

\subsection {Gates with Common Targets}
Karnaugh map (\emph{Kmap}) has been extensively used to simplify small irreversible circuits in the past. On the other hand, the available optimization methods for reversible circuits, used pre-defined patterns (e.g., \cite{MaslovTCAD08}) or well-developed data structures (e.g., \cite{Prasad06}) or heuristics (e.g., \cite{SaeediICCAD07}) to improve the cost of synthesized circuits.

The behavior of reversible gates differs from that of the irreversible ones (i.e., NAND, NOR, etc.) significantly. Hence, applying Kmap for the synthesis of a reversible function (i.e., from specification to circuit) may introduce some difficulties due to different gate types. In \cite{WangNANO03} a synthesis method was proposed which used a modified version of Kmap for the synthesis purpose. In this paper, we use Kmap for the optimization of sub-circuits with common targets that can be used to simplify the results of any other synthesis method.

A C$^{n-1}$NOT gate can be represented by a Boolean expression with $n-1$ inputs and one output where gate controls act as the inputs and the target behaves as the output. Hence, a C$^{n-1}$NOT gate can fill one cell of a Kmap of size $n$ (i.e., $n-1$ inputs, one output). In order to extract the simplified circuit, a Kmap-based cell grouping approach similar to the one used in irreversible logic is applied. Method \ref{Alg:1} discusses the method.

\newcounter{ale}

\newcommand{\abc}{\item[\alph{ale})]\stepcounter{ale}}

\newenvironment{liste}{\begin{itemize}}{\end{itemize}}
\newcommand{\aliste}{\begin{liste} \setcounter{ale}{1}}
\newcommand{\zliste}{\end{liste}}

\newenvironment{abcliste}{\aliste}{\zliste}

\begin{algorithm} \label{Alg:1}
Assume that the Kmap of a function is given. In order to extract the circuit from the given Kmap, the following rules should be followed:
\begin{abcliste}
	\abc All cells with the value 1 should be covered in at least one group.
	\abc The size of each group should be 2$^p$ for $p\geq0$.
	\abc Each cell with the value 1 can be used in an odd number of groups.
	\abc Each empty cell (i.e., a cell with the value 0) can be used in an even number of groups.
	\abc The minimum number of groups should be generated.
	\abc The maximum size for each group should be explored.
\end{abcliste}
\end{algorithm}

Cases (a) and (b) come from the Kmap for irreversible logic. Regarding (c) and (d), since each cell illustrates one C$^{n-1}$NOT gate, the odd (even) number of groups means the odd (even) number of consecutive C$^{n-1}$NOT gates. Applying the deletion rule reveals the proposition. The number of groups denotes the number of gates; hence the case (e) deals to the minimum number of gates. In addition, group size affects the number of control lines of the generated gate. To reduce the number of controls, the maximum size for each group should be explored (case (f)).

According to the case (d) of Method \ref{Alg:1}, empty cells in the Kmap for reversible circuits can also be covered in a group of cells with the values of 1. This characterization differs from the Kmap of irreversible logic. It is worthwhile to note that finding a realization for a given Kmap may not be unique.

\begin{theorem} \label{Lemma:4}
Consider a Kmap grouped by applying Method \ref{Alg:1}. Each group defines a gate with $n-p$ controls where $n$ is the sub-circuit size and $2^p$ is the group size.
\end{theorem}

A group with size 1 is equal to a gate with $n-1$ controls. Similarly, a group with size of 2 deletes one control to construct a gate with $n-2$ controls. Repeating this process prove the theorem.

\begin{definition}
\emph{Common-Target Rule} (CTR): Each reversible sub-circuit of size $n$ with common targets can be optimized by applying Method \ref{Alg:1}.
\end{definition}

Due to the lack of a Kmap with one input, CTR cannot be applied to simplify a 2-input sub-circuit. In order to model this case, the restricted CTR (i.e., R-CTR) is defined as follows:

\begin{definition}
\emph{Restricted CTR} (R-CTR): A positive-control CNOT near a negative-control CNOT on the same target is equivalent to a NOT applied on the same target. A negative-control CNOT immediately before or after a NOT on the same target is equivalent to a positive-control CNOT on the target. A negative control CNOT can be simplified to a positive control CNOT followed by a NOT, all gates on the same targets.
\end{definition}

\subsection {Examples}
The following examples describe the proposed approach in more detail. It is worth noting that none of the examples, except Example \ref{Ex:6} and Example \ref{Ex:7}, can be simplified by using the previously published optimization methods \cite{MillerDAC03, MDM:2005, Prasad06}. Since the proposed approach can use negative controls to reduce the quantum cost, it outperforms the available methods. As shown in the experimental results section, the costs of the best-reported benchmark circuits can be improved by our method in some cases.

\begin{figure}[t]
	\centering
		\includegraphics[scale=0.6]{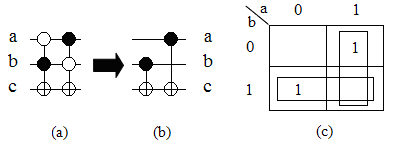}
		\caption{The figure of Example \ref{Ex:1}.
		}
	\label{Fig:6}
\end{figure}

\begin{example} \label{Ex:1}
A circuit with three inputs is shown in Fig. \ref{Fig:6}-a. It can be verified that the quantum cost for this circuit is 10. Since the target lines of both gates are identical, a Kmap of size 2 can be used for the optimization (Fig. \ref{Fig:6}-c). According to Fig. \ref{Fig:6}-c, two groups are found where their sizes are 2$^1$=2. It means that the optimized circuit has two gates (i.e., two groups) and each gate has one control as depicted in Fig. \ref{Fig:6}-b. It can be verified that the quantum cost for the optimized circuit is 2. Note that cells with the values 1 were grouped once and empty cells were grouped twice.
\end{example}

\begin{figure}[t]
	\centering
		\includegraphics[scale=0.45]{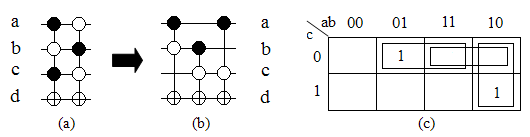}
		\caption{The figure of Example \ref{Ex:2}.
		}
	\label{Fig:7}
\end{figure}

\begin{example} \label{Ex:2}
A reversible circuit of size 4 is shown in Fig. \ref{Fig:7}-a. The optimized circuit and its Kmap are shown in Fig. \ref{Fig:7}-b. and Fig. \ref{Fig:7}-c, respectively. It can be verified that the quantum costs for the circuit before and after the optimization are 26 and 15, respectively.
\end{example}

\begin{figure}[t]
	\centering
		\includegraphics[scale=0.6]{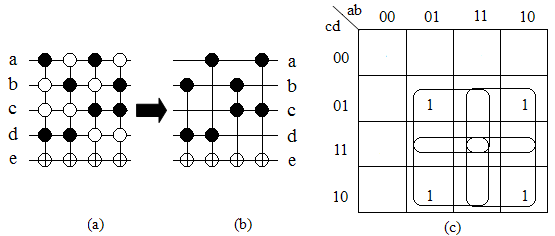}
		\caption{The figure of Example \ref{Ex:3}.
		}
	\label{Fig:8}
\end{figure}

\begin{example} \label{Ex:3}
A reversible circuit of size 5 is shown in Fig. \ref{Fig:8}-a. The optimized circuit and its Kmap are shown in Fig. \ref{Fig:8}-b and Fig. \ref{Fig:8}-c, respectively. It can be verified that the quantum costs for the circuit before and after the optimization are 116 and 20, respectively.
\end{example}

As another example, consider Example \ref{Ex:4} where a 4-input circuit is given. In this case, cells with the values 1 fill almost all cells (for a Kmap with $M$ cells and $m$ 1-cells $m\geq \left\lceil M/2\right\rceil$). In those cases, inverting the Kmap (using zero instead of one) and putting a NOT gate may help. See the following example for more detail.

\begin{figure}[t]
	\centering
		\includegraphics[scale=0.5]{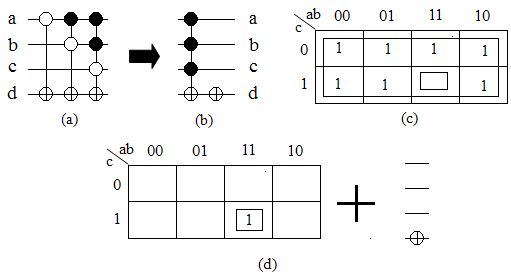}
		\caption{The figure of Example \ref{Ex:4}.
		}
	\label{Fig:9}
\end{figure}

\begin{example} \label{Ex:4}
A circuit of size 4 is shown in Fig. \ref{Fig:9}-a. Based on CTR, \ref{Fig:9}-b shows the realized circuit. Number of cells with the values 1 in more than four (i.e., 8/2=4) as illustrated in Fig. \ref{Fig:9}-c. In Fig. \ref{Fig:9}-d the inverse Kmap is shown and used which leads to a C$^3$NOT followed by a NOT gate.
\end{example}

\begin{figure}[t]
	\centering
		\includegraphics[scale=0.5]{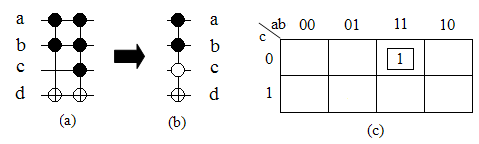}
		\caption{The figure of Example \ref{Ex:5}. }
	\label{Fig:10}
\end{figure}

\begin{example} \label{Ex:5}
A 4-input circuit with only positive controls is shown in Fig. \ref{Fig:10}-a. By applying CTR on this circuit, \ref{Fig:10}-b is resulted. The respective Kmap is shown in Fig. \ref{Fig:10}-c. Note that the term `$abc$' is used twice in the circuit and it can be canceled. The quantum cost is improved from 18 to 13.
\end{example}

Applying GPR (or PR) on some circuits provides opportunities to improve the cost. See the following examples for more detail.

\begin{figure}[t]
	\centering
		\includegraphics[scale=0.65]{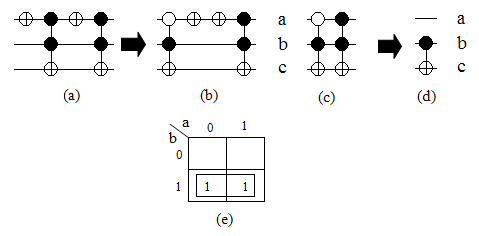}
		\caption{The figure of Example \ref{Ex:6}.
}
	\label{Fig:11}
\end{figure}

\begin{example} \label{Ex:6}
A reversible circuit of size 3 is shown in Fig. \ref{Fig:11}-a. By applying PR on the circuit shown in \ref{Fig:11}-a, the circuit depicted in Fig. \ref{Fig:11}-b is resulted. It can be verified that two NOTs in \ref{Fig:11}-b can be canceled (see Fig. \ref{Fig:11}-c). Now, by applying CTR on the circuit of Fig. \ref{Fig:11}-c, the circuit given in Fig. \ref{Fig:11}-d is provided. In this example, the quantum cost is improved from 12 to 1.
\end{example}

\begin{figure}[t]
	\centering
		\includegraphics[scale=0.4]{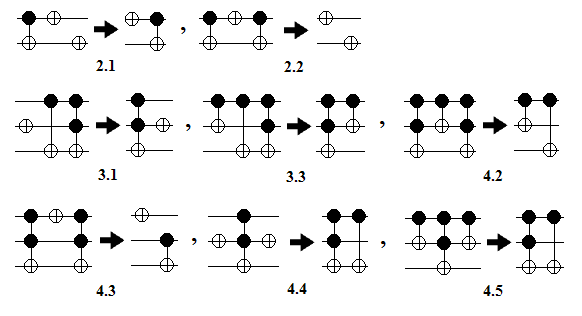}
		\caption{Some templates with two or three inputs introduced in \cite{MillerDAC03}}
	\label{Fig:12}
\end{figure}

\begin{example} \label{Ex:7}
Fig. \ref{Fig:12} shows several templates with 2 and 3 inputs introduced in \cite{MillerDAC03}. In this example, we show how such templates and their generalized forms, resulted by inserting identical controls for all gates, can be obtained from the proposed approach. It can be verified that applying the proposed approach as listed below leads to the same results of \cite{MillerDAC03} (cases 2.1, 2.2, 3.1, 3.3, 4.2, 4.3) or better ones (cases 4.4, 4.5).
\begin{itemize}
	\item 2.1: PR, R-CTR
	\item 2.2: PR, R-CTR
	\item 3.1: CTR, PR
	\item 3.3: CTR, GPR
	\item 4.2: GPR, CTR
	\item 4.3: PR, CTR
	\item 4.4: PR (improved, cost=5)
	\item 4.5: GPR (improved, cost=5)
\end{itemize}
\end{example}

\section{Experimental Results} \label{sec:exper_res}

The proposed rule-based optimization algorithm was implemented in C++ and all of the experiments were done on an Intel Pentium IV 2.2GHz computer with 2GB memory. In order to evaluate the proposed approach, we used the algorithm of \cite{SaeediICCAD07} which used both positive and negative control Toffoli gates. In addition, in \cite{SaeediICCAD07} the truth table of a given reversible function is treated column-wise. Therefore, this algorithm produces sub-circuits with common-target gates in many situations. To compare our results, we used the same set of circuits as used in \cite{SaeediICCAD07}. The synthesis results are shown in Table \ref{table:exper2}. It can be seen that the proposed approach can be used to reduce the quantum cost of a given circuit with both negative and positive controls in some cases significantly.

\begin{table}[t]
\caption{Simplification of the circuits from \cite{SaeediICCAD07}. Runtime for each circuit is less than one second.}
\label{table:exper2}
\centering
\scriptsize

\begin{tabular}{|c|c|l|cc|c|}
\hline
  & & & \multicolumn{2}{c|}{Quantum cost} & Imp. \\
 \# Circuit& $n$ & Specification & \cite{SaeediICCAD07} & Ours & \%\\
\hline
\hline
1 & 3 & (1,0,3,2,5,7,4,6) & 18 & 17 & 5.5 \\
\hline
2 & 3 & (7,0,1,2,3,4,5,6) & 7 & 7 & 0 \\
\hline
3 & 3 & (0,1,2,3,4,6,5,7) & 15 & 15 & 0 \\
\hline
4 & 3 & (0,1,2,4,3,5,6,7) & 27 & 27 & 0 \\
\hline
5 & 4 & (0,1,2,3,4,5,6,8,7,9,10,11,12,13,14,15) & 195 & 131 & 33 \\
\hline
6 & 3 & (1,2,3,4,5,6,7,0) & 10 & 7 & 30 \\
\hline
7 & 4 & (1,2,3,4,5,6,7,8,9,10,11,12,13,14,15,0) & 25 & 20 & 20 \\
\hline
8 & 4 & (0,7,6,9,4,11,10,13,8,15,14,1,12,3,2,5) & 12 & 12 & 0 \\
\hline
9 & 3 & (3,6,2,5,7,1,0,4) & 32 & 29 & 9 \\
\hline
10 & 3 & (1,2,7,5,6,3,0,4) & 35 & 26 & 26 \\
\hline
11 & 3 & (4,3,0,2,7,5,6,1) & 37 & 29 & 22 \\
\hline
12 & 3 & (7,5,2,4,6,1,0,3) & 28 & 19 & 32 \\
\hline
13 & 4 & (6,2,14,13,3,11,10,7,0,5,8,1,15,12,4,9) & 214 & 136 & 36 \\
\hline

\end{tabular}
\end{table}

To further analyze the method, we used the available reversible synthesized benchmarks \cite{MaslovSite}. To this end, the best-reported synthesized circuits were selected \cite{MaslovTODAES07}. Quantum cost is used in all comparisons. For the approach of \cite{MaslovTODAES07}, the synthesis algorithm, the template matching method, the random and exhaustive driver algorithms were applied sequentially to improve synthesis costs. A random driver performs several iterations where at each iteration, a number of random subnetworks are re-synthesized and the best circuit is forwarded to the next iteration. In addition, the exhaustive driver tries all possible subnetworks with at least 5 gates of a given network. Bidirectional and quantum cost reduction modes were also applied.

\begin{figure}[t]
	\centering
		\includegraphics[scale=0.5]{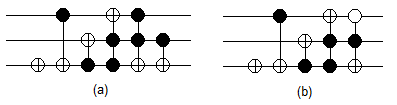}
		\caption{Realization of the 3\_17 benchmark, (a) the best-reported circuit \cite{MaslovTODAES07}, (b) the improved circuit.}
	\label{Fig:13}
\end{figure}

\begin{figure*}[t]
	\centering
		\includegraphics[scale=0.45]{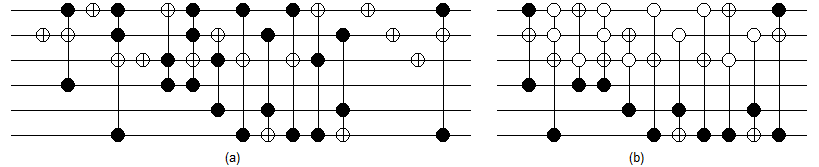}
		\caption{Realization of the mod5adder benchmark, (a) the best-reported circuit \cite{MaslovTODAES07}, (b) the improved circuit.}
	\label{Fig:14}
\end{figure*}

The results of applying the proposed approach on the results of \cite{MaslovTODAES07} are shown in Table \ref{table:exper1}. In this table, quantum costs for both methods are compared. While the results of \cite{MaslovTODAES07} were improved by using different scenarios (i.e., template matching, random driver, exhaustive driver, bidirectional mode, quantum cost reduction mode), the fact that the best-reported synthesized results can be improved by using Toffoli gates with negative controls as done in the proposed approach is very interesting. The application of the proposed approach on non-reported benchmark circuits had no effect on their quantum costs and ignored in Table \ref{table:exper1} to save space. Fig. \ref{Fig:13} and Fig. \ref{Fig:14} illustrate the synthesized circuits of \cite{MaslovTODAES07} and ours for two benchmarks. The runtime of the proposed method is less than a second for each attempted circuit.

\begin{table}[t]
\caption{Improving the best-reported costs of some available benchmarks. Runtime for each circuit is less than one second.}
\label{table:exper1}
\centering
\scriptsize
\begin{tabular}{|c|c|cc|c|}
\hline
 Benchmark & & \multicolumn{2}{c|}{Quantum cost} \\
 Function & $n$ & \cite{MaslovTODAES07} & Ours \\
\hline
\hline
3\_17 & 3 & 14 & 13	\\
\hline
4\_49 & 4 & 32 & 30 \\
\hline
t-add-8 & 24 & 322 & 314 \\
\hline
mod5adder & 6 & 77 & 71 \\
\hline
rd53 & 7 & 65 & 62 \\
\hline
hwb5 & 5 & 104 & 101 \\
\hline
hwb6 & 6 & 142 & 140 \\
\hline
hwb7 & 7 & 2521 & 2516 \\
\hline
hwb8 & 8 & 6709 & 6687 \\
\hline
hwb9 & 9 & 20224 & 20207 \\
\hline
hwb10 & 10 & 52245 & 52225 \\
\hline
hwb11 & 11 & 121840 & 121830 \\
\hline

\end{tabular}
\end{table}

\section {Conclusion} \label{sec:conc}
In this paper, an optimization approach for reversible circuits was proposed which applies a set of rules to improve the quantum cost of a given circuit. By employing both negative and positive control Toffoli gates during the optimization, it was shown that there is some room for improvement in the results of the available synthesis algorithms. To this end, we proposed a rule to reduce NOT gates of a given circuits. Next, a Karnaugh map-based optimization method was presented to optimize sub-circuits with common-target gates. To evaluate the proposed approach, one of the recent synthesis algorithms which used both negative and positive controls was selected. It has been shown that the proposed approach can reduce the quantum cost for the attempted circuits by up to 36\%. Moreover, the experiments showed that our method could improve the best-reported circuits in some cases.

\footnotesize


\end{document}